\documentclass[aps,pra,showpacs,preprint]{revtex4}
\usepackage{graphicx}
\usepackage{amsmath,amssymb,amsfonts}
\usepackage{dcolumn}
\usepackage[english]{babel}
\usepackage{epsfig}

\begin{document}

\title{Representation of the three-body Coulomb Green's function in parabolic coordinates:
 paths of integration}
\author{ S. A. Zaytsev}
\email[E-mail: ]{zaytsev@fizika.khstu.ru} \affiliation{Pacific
National University, Khabarovsk, 680035, Russia}

\begin{abstract}
The possibility is discussed of using straight-line paths of
integration in computing the integral representation of the
three-body Coulomb Green's function. In our numerical examples two
different integration contours are considered. It is demonstrated
that only one of these straight-line paths provides that the
integral representation is valid.
\end{abstract}
 \maketitle

\section{Introduction}
In two previous papers \cite{JP1,JP2} the method was introduced as a
new approach for solution of the three-body continuum problem using
infinite set of $L^2$ parabolic Sturmian basis functions for the
wave function of the system. The goal of these papers has been the
construction of exact analytic matrix elements of the three-body
Coulomb Green's function. The corresponding six-dimensional
resolvent operator has been expressed as a convolution integral of
three two-dimensional Green's function. In this paper we wish to
learn how to choose appropriate straight-line paths of integration
which provide that the integral representation is valid.

Below we outline how the Schr\"{o}dinger equation for a three-body
Coulomb system is transformed into a Lippmann-Schwinger equation in
terms of generalized parabolic coordinates.

The Schr\"{o}dinger equation for three particles with masses $m_1$,
$m_1$, $m_3$ and charges $Z_1$, $Z_2$, $Z_3$ is
\begin{equation}\label{SE}
    \left[-\frac{1}{2\mu_{12}}\Delta_{\bf R}
    -\frac{1}{2\mu_{3}}\Delta_{\bf r}+\frac{Z_1Z_2}{r_{12}}
    +\frac{Z_2Z_3}{r_{23}}+\frac{Z_1Z_3}{r_{13}}\right]\Psi=E\Psi,
\end{equation}
where ${\bf R}$ and ${\bf r}$ are the Jacobi vectors
\begin{equation}\label{Rr}
{\bf R}={\bf r}_1-{\bf r}_2, \quad {\bf r}={\bf r}_3-\frac{m_1{\bf
r}_1+m_2{\bf r}_2}{m_1+m_2},
\end{equation}
${\bf r}_{ls}={\bf r}_l-{\bf r}_s$, $r_{ls}=\left|{\bf r}_{ls}
\right|$, $\mu_{12}$ and $\mu_{3}$ are the reduced masses
\begin{equation}\label{mass}
    \mu_{12}=\frac{m_1m_2}{m_1+m_2}, \quad
    \mu_{3}=\frac{m_3\left(m_1+m_2\right)}{m_1+m_2+m_3}.
\end{equation}
The ansatz
\begin{equation}\label{PsiO}
    \Psi=e^{i({\bf K}\cdot{\bf R}+{\bf k}\cdot{\bf r})}\overline{\Psi}
\end{equation}
removes the eigenenergy $E=\frac{1}{2\mu_{12}}\,{\bf
K}^2+\frac{1}{2\mu_{3}}\,{\bf k}^2$ giving the equation for
$\overline{\Psi}$
\begin{equation}\label{SEO}
    \left[-\frac{1}{2\mu_{12}}\,\Delta_{\bf R}
    -\frac{1}{2\mu_{3}}\,\Delta_{\bf r}
    -\frac{i}{\mu_{12}}\,{\bf K}\cdot \nabla_{\bf R}
    -\frac{i}{\mu_{3}}\,{\bf k}\cdot \nabla_{\bf r}
    +\frac{Z_1Z_2}{r_{12}}+\frac{Z_2Z_3}{r_{23}}
    +\frac{Z_1Z_3}{r_{13}}\right]\overline{\Psi}=0.
\end{equation}
Then, the operator in the square braces is expressed in terms of the
generalized parabolic coordinates \cite{Klar}
\begin{equation}\label{pc}
  \begin{array}{c}
  \xi_1=r_{23}+\hat{\bf k}_{23}\cdot{\bf r}_{23}, \quad
  \eta_1=r_{23}-\hat{\bf k}_{23}\cdot{\bf r}_{23},\\
  \xi_2=r_{13}+\hat{\bf k}_{13}\cdot{\bf r}_{13}, \quad
  \eta_2=r_{13}-\hat{\bf k}_{13}\cdot{\bf r}_{13},\\
  \xi_3=r_{12}+\hat{\bf k}_{12}\cdot{\bf r}_{12}, \quad
  \eta_3=r_{12}-\hat{\bf k}_{12}\cdot{\bf r}_{12},\\
  \end{array}
\end{equation}
where ${\bf k}_{ls}=\frac{{\bf k}_l m_s-{\bf k}_s m_l}{m_l+m_s}$ is
the relative momentum, $\hat{\bf k}_{ls}=\frac{{\bf
k}_{ls}}{k_{ls}}$, $k_{ls}=\left|{\bf k}_{ls}\right|$. In the
resulting equation
\begin{equation}\label{PSEO}
    \left[\hat{D}_0+\hat{D}_1 \right]\overline{\Psi}=0
\end{equation}
the first operator is given by
\begin{equation}\label{D1}
    \hat{D}_0=\sum \limits_{j=1}^{3}\frac{1}{\mu_{ls}\left(\xi_j+\eta_j \right)}
    \left[ \hat{h}_{\xi_j}+\hat{h}_{\eta_j}+2k_{ls} t_{ls}\right],
\end{equation}
for $j\neq l,\,s$ and $l<s$. Here $t_{ls}=\frac{Z_l Z_s
\mu_{ls}}{k_{ls}}$, $\mu_{ls}=\frac{m_l m_s}{m_l+m_s}$; the
one-dimensional operators $\hat{h}_{\xi_j}$ and $\hat{h}_{\eta_j}$
are
\begin{equation}\label{hxiheta}
    \hat{h}_{\xi_j}=-2\left(\frac{\partial}{\partial \xi_j}
    \xi_j\frac{\partial}{\partial \xi_j}+i k_{ls}\xi_j\frac{\partial}{\partial \xi_j} \right), \quad
    \hat{h}_{\eta_j}=-2\left(\frac{\partial}{\partial \eta_j}
    \eta_j\frac{\partial}{\partial \eta_j}-i k_{ls}\eta_j\frac{\partial}{\partial \eta_j}
    \right).
\end{equation}
$\hat{D}_0$ is the leading term which provides a three-body
continuum wave function that satisfies exact asymptotic boundary
conditions for Coulomb systems, when the three particles are far
away from each other \cite{Klar}. In turn, the operator $\hat{D}_1$
(which contains the non-orthogonal part of the kinetic energy
operator) is regarded as a small perturbation which does not violate
the boundary conditions.

The best known approximate solution to the equation (\ref{PSEO}),
the so-called C3 model \cite{Klar,C31,C32,C33}, is obtained by
neglecting of $\hat{D}_1$. Many improvements to the C3 model have
been developed by considering in some approximate way of the
neglected terms of the kinetic energy (see, e. g., \cite{C3mod} and
references therein). In our approach the wave function
$\overline{\Psi}$ is obtained by solving an equivalent
Lippman-Schwinger integral equation. We multiply (\ref{PSEO}) by
$\prod \limits _{j=1}^3\mu_{ls}\left(\xi_j+\eta_j \right)$ from the
left before the transformation of (\ref{PSEO}) into an integral
equation. Thus, in the resulting equation
\begin{equation}\label{LSE}
    \overline{\Psi}=\overline{\Psi}^{(0)}-\hat{\mathfrak{G}}\hat{V}\overline{\Psi}
\end{equation}
$\hat{\mathfrak{G}}$ plays the role of Green's function operator
which is formally inverse to the six-dimensional operator
$\hat{\mathfrak{h}}$ given by
\begin{equation}\label{h}
 \begin{array}{c}
\hat{\mathfrak{h}}\equiv\prod \limits
_{j=1}^3\mu_{ls}\left(\xi_j+\eta_j
\right)\hat{D}_0=\mu_{13}\left(\xi_2+\eta_2
\right)\mu_{12}\left(\xi_3+\eta_3 \right)\hat{\mathfrak{h}}_1\\[3mm]
+\mu_{23}\left(\xi_1+\eta_1 \right)\mu_{12}\left(\xi_3+\eta_3
\right)\hat{\mathfrak{h}}_2 +\mu_{23}\left(\xi_1+\eta_1
\right)\mu_{13}\left(\xi_2+\eta_2 \right)\hat{\mathfrak{h}}_3,\\
 \end{array}
\end{equation}
\begin{equation}\label{hj}
\hat{\mathfrak{h}}_j=\hat{h}_{\xi_j}+\hat{h}_{\eta_j}+2k_{ls}t_{ls}.
\end{equation}
The ``potential'' $\hat{V}$ is defined as
\begin{equation}\label{V}
\hat{V}=\prod \limits _{j=1}^3\mu_{ls}\left(\xi_j+\eta_j
\right)\hat{D}_1.
\end{equation}
The inhomogeneous term $\overline{\Psi}^{(0)}$ of Eq. (\ref{LSE})
can be taken as the wave function of the C3 model, i. e. expressed
in terms of a product of three Coulomb waves.

It has been suggested in \cite{JP1,JP2} to treat the equation within
the context of $L^2$ parabolic Sturmian basis set \cite{Ojha1}
\begin{equation}\label{B6}
    \left|\mathfrak{N} \right>=\prod \limits _{j=1}^{3} \phi_{n_j\,m_j}\left(\xi_j,\,\eta_j
    \right),
\end{equation}
\begin{equation}\label{B2}
\phi_{n_j\,m_j}\left(\xi_j,\,\eta_j \right)=\psi_{n_j}\left(\xi_j
\right)\psi_{m_j}\left(\eta_j\right),
\end{equation}
\begin{equation}\label{B1}
    \psi_{n}\left(x\right)=\sqrt{2b}e^{-bx}L_{n}(2bx),
\end{equation}
where $b$ is the scaling parameter. A solution $\overline{\Psi}$ of
the Lippman-Schwinger equation (\ref{LSE}) is expanded in basis
(\ref{B6}) as
\begin{equation}\label{PsiE}
\overline{\Psi}=\sum \limits _{\mathfrak{N}}
a_{\mathfrak{N}}\left|\mathfrak{N} \right>.
\end{equation}
The discrete analog of the Lippman-Schwinger equation is obtained by
putting (\ref{LSE}) in the basis set $(\ref{B6})$. This yields
\begin{equation}\label{DALSE}
    \underline{a}=\underline{a}^{(0)}-\underline{\mathfrak{G}}\,\underline{V}\,\underline{a},
\end{equation}
where $\underline{\mathfrak{G}}$ and $\underline{V}$ are the
operators $\hat{\mathfrak{G}}$ and $\hat{V}$ matrix representations
in basis (\ref{B6}), $\underline{a}$ and $\underline{a}^{(0)}$ are
the coefficient vectors of $\overline{\Psi}$ and
$\overline{\Psi}^{(0)}$ respectively.

It has been shown in our previous paper \cite{JP2} that the matrix
$\underline{\mathfrak{G}}$ can be represented in the form of a
convolution integral
\begin{equation}\label{G6}
 \begin{array}{c}
    \underline{\mathfrak{G}}^{(\pm)}=\frac{\aleph}{(2\pi i)^2}\int \limits_{\mathcal{C}^{(1)}}\int
    \limits_{\mathcal{C}^{(2)}}\frac{d\mathcal{E}_1}{\mu_{23}} \frac{d\mathcal{E}_2}{\mu_{13}}\, {\bf
    G}^{(\pm)}\left(t_{23};\; \mathcal{E}_1 \right)\otimes {\bf
    G}^{(\pm)}\left(t_{13};\; \mathcal{E}_2 \right)\\[3mm]
    \otimes {\bf G}^{(\pm)}\left(t_{12};\; \mathcal{E}_3=\frac{k_{12}^2}{2}
    +\frac{\mu_{12}}{\mu_{23}}\left(\frac{k_{23}^2}{2}-\mathcal{E}_1 \right)
    +\frac{\mu_{12}}{\mu_{13}}\left(\frac{k_{13}^2}{2}-\mathcal{E}_2 \right)
    \right).\\
 \end{array}
\end{equation}
Here ${\bf G}^{(\pm)}\left(t_{ls},\, \mathcal{E}_j \right)$ is the
matrix which is inverse of the two-dimensional operator
$\left[\hat{\mathfrak{h}}_j+\left(\frac{k_{ls}^2}{2}-\mathcal{E}_j\right)
\left(\xi_j+\eta_j \right) \right]$ matrix representation in the
basis (\ref{B2}), i. e.
\begin{equation}\label{hg2}
    \left[{\bf h}_j+\left(\frac{k_{ls}^2}{2}-\mathcal{E}_j\right){\bf
    Q}_j\right]{\bf G}^{(\pm)}\left(t_{ls},\, \mathcal{E}_j
    \right)={\bf I}_j.
\end{equation}
Here the matrix ${\bf h}_j$ of the operator $\hat{\mathfrak{h}}_j$
(\ref{hj}) is expressed in terms of the one-dimensional operators
(\ref{hxiheta}) matrices:
\begin{equation}\label{hjm}
    {\bf h}_j={\bf h}_{\xi_j}\otimes{\bf I}_{\eta_j}+{\bf I}_{\xi_j}\otimes{\bf
    h}_{\eta_j}+2k_{ls}t_{ls}{\bf I}_j.
\end{equation}
In Eqs. (\ref{hg2}) and (\ref{hjm}) ${\bf I}_{\xi_j}$, ${\bf
I}_{\eta_j}$ and ${\bf I}_j={\bf I}_{\xi_j}\otimes{\bf I}_{\eta_j}$
are the unit matrices. ${\bf Q}_j={\bf Q}_{\xi_j}\otimes{\bf
I}_{\eta_j}+{\bf I}_{\xi_j}\otimes{\bf Q}_{\eta_j}$, where ${\bf
Q}_{\xi_j}$ and ${\bf Q}_{\eta_j}$ are the matrices of $\xi_j$ and
$\eta_j$ in basis (\ref{B1}), respectively.

In this paper we make use of matrices ${\bf G}^{(\pm)}\left(t,\,
\mathcal{E}\right)$ (\ref{Gpm}) which are more symmetric (in $\xi$
and $\eta$) than that obtained in our previous work \cite{JP2}. For
simplicity of the notation, we omit indices for a while. The new
matrix, e. g., ${\bf G}^{(+)}\left(t,\, \mathcal{E}\right)$ also
obeys the completeness relation
\begin{equation}\label{CRS}
    \frac{1}{2\pi i}\int \limits _{\mathcal{C}} d\mathcal{E}\,{\bf
    G}^{(+)}(t;\;\mathcal{E})=\left[{\bf Q}_{\xi}\otimes{\bf
I}_{\eta}+{\bf I}_{\xi}\otimes{\bf Q}_{\eta}\right]^{-1},
\end{equation}
established in \cite{JP2}. Here $\mathcal{C}$ is a contour
originating at $\mathcal{E}=\infty$, below the positive real axis
rounding the lowest bound state $\mathcal{E}_1=-\frac{\left(k
t\right)^2}{2}$ for $t<0$ (or the origin for $t>0$), and then
heading back to $\mathcal{E}=\infty$ --- this time staying above the
cut (see Fig.~1). The integration contours of the convolution
integral (\ref{G6}) $\mathcal{C}^{(1,2)}$ are similar to the contour
$\mathcal{C}$ (see e. g. \cite{Faddeev}). However, despite the known
paths of integration the representation (\ref{G6}) poses several
difficulties in practical applications, the most serious of which is
that one cannot trace crossing the cut along the positive real axis
in $\mathcal{E}_3$-plane during integration over
$\mathcal{E}_{1,2}$. The (numerical) evaluation of the integral
(\ref{G6}) can be simplified considerably by using straight lines as
paths of integration. In this paper we wish to learn how to choose
appropriate straight-line paths $\mathcal{C}^{(1)}$ and
$\mathcal{C}^{(2)}$ for which the integral representation (\ref{G6})
is valid. Unfortunately, the contour integrals of interest cannot be
treated analytically, so we must resort to numerical experiments.

The numerical examples presented in Section III show that the
contour $\mathcal{C}$ in (\ref{CRS}) could be deformed so that it
becomes the disconnected pair of straight lines. The value of the
integral over each of these straight-line paths is half the value of
the contour integral (\ref{CRS}). In this section we consider two
kinds of straight-line paths. In Section IV based upon the numerical
results obtained for double integrals which arise from the matrix
product $\underline{\mathfrak{h}}\,\underline{\mathfrak{G}}^{(+)}$,
we find the straight-line contours $\mathcal{C}^{(1,\,2)}$ providing
a non-zero integral representation (\ref{G6}). Section V contains a
brief discussion of the overall results. For completeness we review
briefly the results of our previous works \cite{JP1,JP2} in the
Appendix.

\section{Preliminaries}
We assume that the relationships between contour integrals obtained
in this work result from the Green's functions properties and are
independent of the base function (\ref{B6}) numbers (that specify
indices of the matrix elements of the operators). Thus in all our
numerical examples (except for the case where the inverse
relationship between $\underline{\mathfrak{h}}$ and
$\underline{\mathfrak{G}}$ is demonstrated) we only use the element
$G^{(+)}_{0,\,0;\; 0,\, 0}\left(t_{ls};\;\mathcal{E}_j\right)\equiv
G^{(+)}_{0}\left(t_{ls};\;\mathcal{E}_j\right)$ of the matrices
${\bf G}^{(+)}\left(t_{ls};\;\mathcal{E}_j\right)$. Notice that the
completeness relation (\ref{CRS}) for $G^{(+)}_0(t;\;\mathcal{E})$
takes the form
\begin{equation}\label{CRS0}
    \frac{1}{2\pi i}\int \limits _{\mathcal{C}} d\mathcal{E}
    \,G^{(+)}_0(t;\;\mathcal{E})=2b.
\end{equation}

We consider two electrons in the Coulomb field of $\mbox{He}^{++}$,
i. e. $Z_1=Z_2=-1$ and $Z_3=2$ (we use atomic units hereafter).
Infinite mass $m_3$ for the nucleus is adopted. Let the two
electrons move in opposite directions with equal energies $E_1=E_2$,
i. e. ${\bf k}_{13}={\bf k}$ and ${\bf k}_{23}=-{\bf k}$. Note that
in this case ${\bf k}_{12}=\frac{1}{2}\left({\bf k}_{13}-{\bf
k}_{23} \right)={\bf k}$ (so that matrix elements of the
``potential'' operator $\hat{V}$ (\ref{V}) could be evaluated
analytically). Further we set the electron energies
$E_1=E_2=25/\mbox{Ry}$ and the scaling parameter $b = 1$.

\section{Straight-line paths}
Clearly, the contour $\mathcal{C}$ can be deformed by moving its
left edge toward minus infinity so that it becomes the disconnected
pair of straight lines ($\mathcal{C}_1$ and $\mathcal{C}_2$ in
Fig.~1) parallel to the real axis. This deformation does not alter
the value of the integral
\begin{equation}\label{vnm}
    v_{n\,m;\;n'\,m'}=\frac{1}{2\pi
    i}\int \limits _{\mathcal{C}}
    d\mathcal{E}\,G^{(+)}_{n\,m;\;n'\,m'}\left(t;\;\mathcal{E}\right).
\end{equation}
Hence $v_{n\,m;\;n'\,m'}$ is divided into two contributions:
\begin{equation}\label{vnmd}
    v_{n\,m;\;n'\,m'}=v^{(1)}_{n\,m;\;n'\,m'}+v^{(2)}_{n\,m;\;n'\,m'},
\end{equation}
where
\begin{equation}\label{vnm1}
    v^{(1)}_{n\,m;\;n'\,m'}=\frac{1}{2\pi
    i}\int \limits _{\mathcal{C}_1}
    d\mathcal{E}\,G^{(+)}_{n\,m;\;n'\,m'}\left(t;\;\mathcal{E}\right),
\end{equation}
\begin{equation}\label{vnm2}
    v^{(2)}_{n\,m;\;n'\,m'}=\frac{1}{2\pi
    i}\int \limits _{\mathcal{C}_2}
    d\mathcal{E}\,G^{(+)}_{n\,m;\;n'\,m'}\left(t;\;\mathcal{E}\right).
\end{equation}
Consider the integral (\ref{vnm1}). Notice that the energy
$\mathcal{E}$ is parametrized by $\mathcal{E}=i y_0+x$ on the
contour $\mathcal{C}_1$. Obviously, the value of
$v^{(1)}_{n\,m;\;n'\,m'}$ is independent of the parameter $y_0$
which specifies the position of $\mathcal{C}_1$ with respect to the
real axis (see Fig.~1). The numerical results for the integrals
\begin{equation}\label{v01}
    v^{(1)}_0=\int \limits _{-\infty}^{\infty}d x\,\frac{1}{2\pi
    i}\,G^{(+)}_0\left(-\frac{2}{k};\;i y_0+x\right)
\end{equation}
with different $y_0$ are presented in Table~1. Note that the
integrand in (\ref{v01}) involves hypergeometric functions ${_2
F_1(a,\, b;\; a+b;\; z)}$. We found that the Gauss continued
fraction \cite{Erdelyi} (rather than the infinite sum (15.3.10) in
\cite{Abramowitz}) provides an efficient approximation for the
hypergeometric functions. The integrals are computed using IMSL
FORTRAN Library routines. The real and imaginary parts of the
integrand identified as $A(x)$ and $B(x)$ are displayed in Fig.~2
for different $y_0$. It is seen in Fig.~2 that $B(x)$ tends to zero
as the parameter $y_0$ increases uniformly with respect to $x$. This
property of $B(x)$ is consistent with the negligible imaginary parts
of the integrals in Table~1. Using this observation, and
(\ref{CRS0}), we conclude that
\begin{equation}\label{v10v20v0}
  v^{(1)}_0=v^{(2)}_0=1=\frac{1}{2}\,v_0.
\end{equation}
Moreover, our numerical computations show that
\begin{equation}\label{v1v2v}
  v^{(1)}_{n\,m;\;n'\,m'}=v^{(2)}_{n\,m;\;n'\,m'}
  =\frac{1}{2}\,v_{n\,m;\;n'\,m'}
\end{equation}
holds for arbitrary $n,\, m, \, n', \, m'$.

Another straight-line path is obtained by rotating the contour
$\mathcal{C}_1$ about some point $x_0$ on the positive real axis
through an angle $\varphi$ in the range $-\pi< \varphi <0$
\cite{Shakeshaft,JP2}. For definiteness, we choose $x_0=\frac{k}{2}$
and $\varphi=-\frac{\pi}{2}$, i. e. $\mathcal{E}=\frac{k}{2}+iy$ on
the resulting contour $\mathcal{C}_3$, shown in Fig.~3. The path
$\mathcal{C}_3$ crosses the cut so that its lower part (depicted in
Fig.~3 by the dashed line) descends into the ``unphysical'' sheet
($-2\pi < \arg(\mathcal{E})<0$). To analytically continue matrix
elements of ${\bf g}^{\xi(+)}$ and ${\bf g}^{\eta(+)}$ in
(\ref{Gpm}) onto the unphysical sheet we use the formulae
(\ref{acfxi}) and (\ref{acfeta}). Note that the numerical result for
the integral
\begin{equation}\label{I1C3}
    v^{(3)}_0=\frac{1}{2\pi}\,\int \limits _{\infty}^{-\infty} dy\,
    G^{(+)}_{0}\left(-\frac{2}{k};\;\frac{k^2}{2}+iy\right)
\end{equation}
presented in Table~1 also satisfies
\begin{equation}\label{v30v0}
  v^{(3)}_0=\frac{1}{2}\,v_0.
\end{equation}
Thus, we find another straight-line path $\mathcal{C}_3$ for which
\begin{equation}\label{vnm3}
    v^{(3)}_{n\,m;\;n'\,m'}=\frac{1}{2\pi
    i}\int \limits _{\mathcal{C}_3}
    d\mathcal{E}\,G^{(+)}_{n\,m;\;n'\,m'}\left(t;\;\mathcal{E}\right)
    =\frac{1}{2}\,v_{n\,m;\;n'\,m'}.
\end{equation}

\section{Double integrals}
Now that we have the relationships (\ref{v1v2v}) and (\ref{vnm3})
between the integrals along the contour $\mathcal{C}$ and the
integrals over the straight-line paths, we can use  $\mathcal{C}_1$
and $\mathcal{C}_3$ in the integral representation (\ref{G6}) of the
Green's function operator. Before proceeding, we consider the matrix
product $\underline{\mathfrak{h}}\,\underline{\mathfrak{G}}^{(+)}$
to determine the normalizing factors $\aleph$ corresponding to the
paths $\mathcal{C}_1$ and $\mathcal{C}_3$. Using the six-dimensional
operator $\hat{\mathfrak{h}}$ (\ref{h}) matrix representation
\begin{equation}\label{hm}
\underline{\mathfrak{h}}=\mu_{13}\mu_{12}{\bf h}_1\otimes{\bf
Q}_2\otimes{\bf Q}_3+\mu_{23}\mu_{12}{\bf Q}_1\otimes{\bf
h}_2\otimes{\bf Q}_3+\mu_{23}\mu_{13}{\bf Q}_1\otimes{\bf
Q}_2\otimes{\bf h}_3,
\end{equation}
(\ref{G6}) and (\ref{hg2}), we have
\begin{equation}\label{h6G6}
    \underline{\mathfrak{h}}\,\underline{\mathfrak{G}}^{(+)}=\aleph\left\{{\bf
    W}_1+{\bf W}_2+{\bf W}_3 \right\},
\end{equation}
where
\begin{equation}\label{MV1}
 \begin{array}{c}
    {\bf W}_1=\frac{1}{(2\pi i)^2}\frac{\mu_{12}}{\mu_{23}}\int \limits _{\mathcal{C}^{(1)}}\int
    \limits _{\mathcal{C}^{(2)}} d\mathcal{E}_1 d\mathcal{E}_2\; {\bf
    I}_1\otimes\left[{\bf Q}_2 {\bf G}^{(+)}\left(t_{13};\; \mathcal{E}_2 \right)
    \right]\\[3mm]
    \otimes\left[{\bf Q}_3 {\bf G}^{(+)}\left(t_{12};\; \frac{k_{12}^2}{2}
    +\frac{\mu_{12}}{\mu_{23}}\left(\frac{k_{23}^2}{2}-\mathcal{E}_1\right)
    +\frac{\mu_{12}}{\mu_{13}}\left(\frac{k_{13}^2}{2}-\mathcal{E}_2\right)
\right) \right],\\
 \end{array}
\end{equation}
\begin{equation}\label{MV2}
 \begin{array}{c}
    {\bf W}_2=\frac{1}{(2\pi i)^2}\frac{\mu_{12}}{\mu_{13}}\int \limits _{\mathcal{C}^{(1)}}\int
    \limits _{\mathcal{C}^{(2)}} d\mathcal{E}_1 d\mathcal{E}_2\; \left[{\bf Q}_1 {\bf G}^{(+)}
\left(t_{23};\; \mathcal{E}_1
     \right)\right]\otimes{\bf I}_2\\[3mm]
    \otimes\left[{\bf Q}_3 {\bf G}^{(+)}\left(t_{12};\; \frac{k_{12}^2}{2}
    +\frac{\mu_{12}}{\mu_{23}}\left(\frac{k_{23}^2}{2}-\mathcal{E}_1\right)
    +\frac{\mu_{12}}{\mu_{13}}\left(\frac{k_{13}^2}{2}-\mathcal{E}_2\right)
\right) \right],\\
 \end{array}
\end{equation}
\begin{equation}\label{MV3}
    {\bf W}_3=\frac{1}{(2\pi i)^2}\int \limits _{\mathcal{C}^{(1)}}\int
    \limits _{\mathcal{C}^{(2)}} d\mathcal{E}_1 d\mathcal{E}_2\;
    \left[{\bf Q}_1 {\bf G}^{(+)}\left(t_{23};\; \mathcal{E}_1 \right)\right]\otimes
    \left[{\bf Q}_2 {\bf G}^{(+)}\left(t_{13};\; \mathcal{E}_2 \right)\right]\otimes{\bf I}_3.
\end{equation}
The matrices $\underline{\mathfrak{h}}$ and
$\underline{\mathfrak{G}}^{(+)}$ must be inverses of each other.
Therefore, the normalizing factor $\aleph$ and the matrices ${\bf
W}_j$, $j=\overline{1,\,3}$ satisfy the condition
\begin{equation}\label{NC}
\aleph\left\{{\bf W}_1+{\bf W}_2+{\bf W}_3 \right\}= {\bf I}.
\end{equation}
If we choose $\mathcal{C}^{(1)}=\mathcal{C}_1$ and
$\mathcal{C}^{(2)}=\mathcal{C}_1$ (or
$\mathcal{C}^{(1)}=\mathcal{C}^{(2)}=\mathcal{C}_3$), then it
follows from (\ref{CRS}) and (\ref{v1v2v}) (or (\ref{vnm3})) that
\begin{equation}\label{V3I}
    {\bf W}_3=\frac{1}{4}\,{\bf I}_1\otimes{\bf I}_2\otimes{\bf I}_3=\frac{1}{4}\,{\bf
    I}.
\end{equation}
Further, we assume that
\begin{equation}\label{V1IV2I}
    {\bf W}_1=\frac{\alpha}{4}\,{\bf I}, \quad  {\bf W}_2=\frac{\beta}{4}\,{\bf I},
\end{equation}
so that the sum $\left\{{\bf W}_1+{\bf W}_2+{\bf W}_3\right\}$ is
proportional to the unit matrix ${\bf I}$. In turn, the constants
$\alpha$ and $\beta$ can be determined by, e. g., the ratios
\begin{equation}\label{ab}
    \alpha=\frac{w_1}{w_3}, \qquad  \beta=\frac{w_2}{w_3},
\end{equation}
where
\begin{equation}\label{v1v2v3}
 \begin{array}{c}
    w_1=\frac{1}{(2\pi i)^2}\frac{\mu_{12}}
    {\mu_{23}}\int \limits _{\mathcal{C}^{(1)}}\int
    \limits _{\mathcal{C}^{(2)}} d\mathcal{E}_1 d\mathcal{E}_2\,
    G^{(+)}_{0}\left(t_{13};\; \mathcal{E}_2 \right)\\[4mm]
    \times G^{(+)}_{0}\left(t_{12};\; \frac{k_{12}^2}{2}
    +\frac{\mu_{12}}{\mu_{23}}\left(\frac{k_{23}^2}{2}-\mathcal{E}_1\right)
    +\frac{\mu_{12}}{\mu_{13}}\left(\frac{k_{13}^2}{2}-\mathcal{E}_2\right)
\right),\\[4mm]
    w_2=\frac{1}{(2\pi i)^2}\frac{\mu_{12}}
    {\mu_{13}}\int \limits _{\mathcal{C}^{(1)}}\int
    \limits _{\mathcal{C}^{(2)}} d\mathcal{E}_1 d\mathcal{E}_2\,
    G^{(+)}_{0}\left(t_{23};\; \mathcal{E}_1 \right)\\[4mm]
    \times G^{(+)}_{0}\left(t_{12};\; \frac{k_{12}^2}{2}
    +\frac{\mu_{12}}{\mu_{23}}\left(\frac{k_{23}^2}{2}-\mathcal{E}_1\right)
    +\frac{\mu_{12}}{\mu_{13}}\left(\frac{k_{13}^2}{2}-\mathcal{E}_2\right)
\right),\\[4mm]
    w_3=\frac{1}{(2\pi i)^2}\int
    \limits _{\mathcal{C}^{(1)}}\int
    \limits _{\mathcal{C}^{(2)}} d\mathcal{E}_1 d\mathcal{E}_2\,
    G^{(+)}_{0}\left(t_{23};\; \mathcal{E}_1 \right)
    G^{(+)}_{0}\left(t_{13};\; \mathcal{E}_2 \right).\\
 \end{array}
\end{equation}
Thereafter, the normalizing factor $\aleph$ is expressed as
\begin{equation}\label{Norm}
\aleph=\frac{4}{1+\alpha+\beta}.
\end{equation}

First we consider the path $\mathcal{C}_1$.

\subsection*{a) $\mathcal{C}^{(1)}=\mathcal{C}^{(2)}=\mathcal{C}_1$}

In this case the energies $\mathcal{E}_1$ and $\mathcal{E}_2$ are
parametrized by $\mathcal{E}_1=iy_0+x_1$ and
$\mathcal{E}_2=iy_0+x_2$ with $y_0=100$. Assuming that the energy
$\mathcal{E}_3=k^2-iy_0-\frac{1}{2}\left(x_1+x_2 \right)$ lies on
the ``physical'' sheet i. e. $0 < \arg(\mathcal{E}_3)<2\pi$, in view
of (\ref{vnm1})-(\ref{v10v20v0}), we obtain that
\begin{equation}\label{v12}
 \begin{array}{c}
    \frac{1}{2\pi i}\int \limits _{-\infty}^{\infty}dx\,G^{(+)}_{0}\left(\frac{1}{2k};\;
k^2-iy_0-\frac{1}{2}x\right)=2\frac{1}{2\pi i}\int \limits
_{-\infty}^{\infty}dx\,G^{(+)}_{0}\left(\frac{1}{2k};\;
k^2-iy_0+x\right)\\[4mm]
 =-2\frac{1}{2\pi i}\int \limits
_{\mathcal{C}_2}d\mathcal{E}\,G^{(+)}_{0}\left(\frac{1}{2k};\; \mathcal{E}\right)=-2v^{(2)}_0=-2v^{(1)}_0.\\
 \end{array}
\end{equation}
Hence, one might expect that the constants $\alpha$ and $\beta$
(\ref{ab}) are negative. Note that in our case the integrals $w_j$
(\ref{v1v2v3}) take the forms
\begin{equation}\label{w1w2w3}
 \begin{array}{c}
    w_1=-\frac{1}{(2\pi)^2}\frac{1}{2}\mathop{\int\int}\limits _{-\infty}^{\infty}
     dx_1 dx_2\,
    G^{(+)}_{0}\left(-\frac{2}{k};\; iy_0+x_2 \right)\,
    G^{(+)}_{0}\left(\frac{1}{2k};\; k^2-iy_0-\frac{1}{2}\left(x_1+x_2 \right)
\right),    \\[4mm]
    w_2=-\frac{1}{(2\pi)^2}\frac{1}{2}\mathop{\int\int}\limits _{-\infty}^{\infty}
     dx_1 dx_2\,
    G^{(+)}_{0}\left(-\frac{2}{k};\; iy_0+x_1 \right) \,
    G^{(+)}_{0}\left(\frac{1}{2k};\;
k^2-iy_0-\frac{1}{2}\left(x_1+x_2\right)\right),
    \\[4mm]
    w_3=-\frac{1}{(2\pi)^2}\mathop{\int\int}\limits _{-\infty}^{\infty}
     dx_1 dx_2\,
    G^{(+)}_{0}\left(-\frac{2}{k};\; iy_0+x_1 \right)\,
    G^{(+)}_{0}\left(-\frac{2}{k};\; iy_0+x_2 \right).\\
 \end{array}
\end{equation}
From the numerical results for the double integrals (\ref{w1w2w3})
presented in Table~1, it follows that $w_1=w_2=-\frac{1}{2}$ and
$w_3=1$, i. e. $\alpha=\beta=-\frac{1}{2}$. Thus, we have
$\alpha+\beta+1=0$, and so the equation (\ref{Norm}) is meaningless.
This outcome is consistent with the numerical result obtained for
the integral in the expression (\ref{G6}) for the diagonal matrix
element of $\left[\underline{\mathfrak{G}}^{(+)}\right]_{0,\,0}$
corresponding to the basis function $\left|0
\right>\equiv\left|n_j=m_j=0,\, j=\overline{1, \,3} \right>$
(\ref{B6}):
\begin{equation}\label{G60}
 \begin{array}{c}
\mathcal{I}_0= \frac{1}{(2\pi i)^2}\mathop{\int\int}\limits
_{-\infty}^{\infty}
     dx_1 dx_2\,
    G^{(+)}_{0}\left(-\frac{2}{k};\; iy_0+x_1 \right)\,
    G^{(+)}_{0}\left(-\frac{2}{k};\; iy_0+x_2 \right)\\[4mm]
    \times G_{0}\left(\frac{1}{2k};\;
k^2-iy_0-\frac{1}{2}\left(x_1+x_2\right)\right),\\
 \end{array}
\end{equation}
which is presented in Table~1. Therefore, we conclude that the
contour $\mathcal{C}_1$ cannot be used in the integral
representation (\ref{G6}).

Now consider the the contour $\mathcal{C}_3$.

\subsection*{b) $\mathcal{C}^{(1)}=\mathcal{C}^{(2)}=\mathcal{C}_3$}

The energies $\mathcal{E}_1$ and $\mathcal{E}_2$ are given by
$\mathcal{E}_1=\frac{k^2}{2}+iy_1$ and
$\mathcal{E}_2=\frac{k^2}{2}+iy_2$ on the contour $\mathcal{C}_3$.
In turn, the energy $\mathcal{E}_3$ is parametrized as
$\mathcal{E}_3=\frac{k^2}{2}-\frac{i}{2}\left(y_1+y_2 \right)$. In
this case the contour integrals (\ref{v1v2v3}) are transformed into
the double integrals
\begin{equation}\label{y1y2y3}
 \begin{array}{c}
    w_1=\frac{1}{(2\pi)^2}\frac{1}{2}\mathop{\int\int}\limits _{+\infty}^{-\infty}
     dy_1 dy_2\,
    G^{(+)}_{0}\left(-\frac{2}{k};\; \frac{k^2}{2}+iy_2 \right)\,
    G^{(+)}_{0}\left(\frac{1}{2k};\;\frac{k^2}{2}-\frac{i}{2}\left(y_1+y_2
\right) \right), \\[4mm]
    w_2=\frac{1}{(2\pi)^2}\frac{1}{2}\mathop{\int\int}\limits _{+\infty}^{-\infty}
     dy_1 dy_2\,
    G^{(+)}_{0}\left(-\frac{2}{k};\;\frac{k^2}{2}+iy_1 \right) \,
    G^{(+)}_{0}\left(\frac{1}{2k};\;\frac{k^2}{2}-\frac{i}{2}\left(y_1+y_2
\right)\right),    \\[4mm]
    w_3=\frac{1}{(2\pi)^2}\mathop{\int\int}\limits _{+\infty}^{-\infty}
     dy_1 dy_2\,
    G^{(+)}_{0}\left(-\frac{2}{k};\;\frac{k^2}{2}+iy_1\right)\,
    G^{(+)}_{0}\left(-\frac{2}{k};\;\frac{k^2}{2}+iy_2 \right).\\
 \end{array}
\end{equation}
Notice that from (\ref{I1C3}) it follows that
\begin{equation}\label{vw3}
    \frac{1}{2\pi}\,\int \limits _{+\infty}^{-\infty} dy\,
    G^{(+)}_{0}\left(\frac{1}{2k};\;\frac{k^2}{2}-\frac{i}{2}y\right)
    =2\left[\frac{1}{2\pi}\,\int \limits _{+\infty}^{-\infty} dy\,
    G^{(+)}_{0}\left(\frac{1}{2k};\;\frac{k^2}{2}+iy\right)\right]=2v^{(3)}_0.
\end{equation}
Therefore, in contrast to the previous case, $w_1$ ($w_2$) would be
expected to have the same sign as $w_3$. From the result for the
numerical evaluations of the integrals (\ref{y1y2y3}), presented in
Table~1., it follows that $w_1=w_2=\frac{3}{2}$ and $w_3=1$. Hence
$\alpha=\beta=\frac{3}{2}$ and $\aleph=1$.

To verify that the straight-line path $\mathcal{C}_3$ does provide
the desired result, we evaluate numerically the matrix elements
\begin{equation}\label{z1z2z3}
 \begin{array}{c}
    \left[\underline{\mathfrak{G}}^{(+)}\right]_{\mathfrak{N},\,\mathfrak{N}'}=
    \frac{1}{(2\pi)^2}\mathop{\int\int}\limits _{+\infty}^{-\infty} dy_1 dy_2\,
    G^{(+)}_{n_1\,m_1;\; n_1'\, m_1'}\left(-\frac{2}{k};\;\frac{k^2}{2}+iy_1\right)\\[4mm]
    \times G^{(+)}_{n_2\,m_2;\; n_2'\,
    m_2'}\left(-\frac{2}{k};\;\frac{k^2}{2}+iy_2\right)\,
    G^{(+)}_{n_3\,m_3;\; n_3'\,
    m_3'}\left(\frac{1}{2k};\;\frac{k^2}{2}-\frac{i}{2}\left(y_1+y_2
    \right)\right)  \\
 \end{array}
\end{equation}
and calculate the matrix product
$\underline{\mathfrak{h}}\,\underline{\mathfrak{G}}^{(+)}$ of finite
size. Notice that the matrix $\underline{\mathfrak{h}}$ (\ref{hm})
is ``tridiagonal'', i. e. for each pair of indices $\left\{n_j,\,
n_j'\right\}$ and $\left\{m_j,\, m_j'\right\}$, $j=\overline{1,\,
3}$, the elements
$\left[\underline{\mathfrak{h}}\right]_{\mathfrak{N}, \,
\mathfrak{N}'}$ vanish unless $\left|n_j-n_j'\right| \leq 1$ and
$\left|m_j-m_j'\right| \leq 1$. Therefore, the minimal rank
$\mathcal{N}_{min}$ of the matrices $\underline{\mathfrak{h}}$ and
$\underline{\mathfrak{G}}$, with which the relation
$\underline{\mathfrak{h}}\underline{\mathfrak{G}}={\bf I}$ could be
verified, is given by $\mathcal{N}_{min}=2^6=64$. Actually, to test
this equality, we must use all the basis functions
$\left|\mathfrak{N}\right>$ (\ref{B6}) with each of the $n_j$ and
$m_j$ taking the value one or zero. Our prime interest here is with
the values of the first row elements
$\left[\underline{\mathfrak{h}}\underline{\mathfrak{G}}^{(+)}\right]_{0,\,
\mathfrak{N}}$ in the matrix
$\underline{\mathfrak{h}}\underline{\mathfrak{G}}^{(+)}$. The
numerical result for the first diagonal element
$\left[\underline{\mathfrak{h}}\underline{\mathfrak{G}}^{(+)}\right]_{0,\,0}$,
presented in Table~1, corresponds to the inverse relationship
between $\underline{\mathfrak{h}}$ and $\underline{\mathfrak{G}}$.
In turn, the remaining (zero) elements of the first row are found to
be of the order of $10^{-7}$.

\section{Conclusion}

In this paper we focus attention on the three-body Coulomb Green's
function operator representation. The development of our method is
based primarily upon the fact that for large particle separation the
Schr\"{o}dinger equation is separable in terms of generalized
parabolic coordinates. Thus, the corresponding six-dimensional
resolvent operator can be expressed as a convolution of three
two-dimensional Greens function. This representation includes
integration along contours, which encircle the spectra of
two-dimensional wave operators. Unfortunately, these (double)
contour integrals are very inconvenient for numerical computation.
Clearly, it is preferable to employ straight-lane paths of
integration. In this paper we demonstrate numerically, with two
simple examples, that use of an appropriate straight-line path of
integration provides a non-zero integral representation of the
Green's function operator.

\section*{Acknowledgments}
The author is grateful to Professor Yu.~V.~Popov for continued
interest in this work and helpful conversations. Additional thanks
are expressed to Dr. V.~Borodulin for his kind hospitality and help.
This work is partially supported by scientific program `Far
East-2008' of the Russian Foundation for Basic Research (regional
grant 08-02-98501).

\appendix{}

\section{Matrix representations of one- and two-dimensional operators}
\subsection{One-dimensional operators}
The matrix representation ${\bf h}_{\xi}$ of the operator
$\hat{\mathfrak{h}}_{\xi}$ (\ref{hxiheta}) in the basis
$\left\{\psi_n(\xi) \right\}_{n=0}^{\infty}$ (\ref{B1}), which is
orthonormal with respect $\xi \in [0,\, \infty)$, is tridiagonal
with nonzero elements
\begin{equation}\label{hxi}
    h^{\xi}_{n,\,n}=b+i k+2 b n,\quad h^{\xi}_{n,\,n-1}=(b-i k)n,
    \quad h^{\xi}_{n,\,n+1}=(b+i k)(n+1).
\end{equation}
In addition, the symmetric matrix ${\bf Q}_{\xi}$ of the operator
$\xi$ in the basis (\ref{B1}) is also tridiagonal:
\begin{equation}\label{Qxieta}
    Q_{n,\, n'}=\left\{
     \begin{array}{lr}
     -\frac{n}{2b}, & n'=n-1,\\
     \frac{2n+1}{2b},& n'=n,\\
     -\frac{n+1}{2b}, & n'=n+1.\\
     \end{array}
    \right.
\end{equation}
Hence, the one-dimensional operator
$\left[\hat{\mathfrak{h}}_{\xi}+2kt+\mu C \xi \right]$ also has the
tridiagonal matrix representation $\left[{\bf h}_{\xi}+2kt{\bf
I}_{\xi}+\mu C {\bf Q}_{\xi} \right]$ (${\bf I}_{\xi}$ is the unit
matrix) in the basis set (\ref{B1}). Fortunately, the inverse of the
matrix $\left[{\bf h}_{\xi}+2kt{\bf I}_{\xi}+\mu C {\bf Q}_{\xi}
\right]$ can be obtained analytically. The elements of the resulting
matrix ${\bf g}^{\xi(\pm)}$ are expressed in terms of well-known
special functions
\begin{eqnarray}\label{gu_pm}
  g^{\xi(+)}_{n_1,\,n_2}(\tau;\; \gamma) &=& \frac{i}{2\gamma}
  \left(\frac{\zeta-1}{\zeta} \right)\frac{\theta^{n_1-n_2}}{\zeta^{n_2}}\,
   p_{n_{<}}(\tau; \; \zeta)\, q_{n_{>}}^{(+)}(\tau; \; \zeta), \\
  g^{\xi(-)}_{n_1,\,n_2}(\tau;\; \gamma) &=& \frac{i}{2\gamma}
  \left(\frac{\zeta-1}{\zeta} \right)\frac{\theta^{n_1-n_2}}{\zeta^{n_2}}\,
   p_{n_{<}}(\tau; \; \zeta)\,\zeta^{n_{>}+1}\, q_{n_{>}}^{(-)}(\tau; \; \zeta),
\end{eqnarray}
where $n_{<}$ is the lesser of $n_1$ and $n_2$, and $n_{>}$ the
greater of the two. Here
\begin{equation}\label{thetalaz}
    \theta=\frac{2b+i(\gamma-k)}{2b-i(\gamma-k)}, \quad
    \lambda=\frac{2b-i(\gamma+k)}{2b+i(\gamma+k)}, \quad
    \zeta=\frac{\lambda}{\theta},
\end{equation}
\begin{equation}\label{tau}
    \tau=\frac{k}{\gamma}\left(t+\frac{i}{2}\right),
\end{equation}
\begin{equation}\label{Eg}
    \mu C=\frac{k^2}{2}-\mathcal{E}, \quad
    \mathcal{E}=\frac{\gamma^2}{2},
\end{equation}
and
\begin{equation}\label{pn}
  p_n(\tau;\; \zeta)=\frac{(-1)^n}{n!}\frac{\Gamma\left(n+\frac{1}{2}-i\tau \right) }
  {\Gamma\left(\frac{1}{2}-i\tau \right) }\;
  {_2F_1\left(-n,\,\frac{1}{2}+i\tau;\; -n+\frac{1}{2}+i\tau;\; \zeta  \right)}
\end{equation}
are the polynomials of degree $n$ in $\tau$ which are orthogonal,
\begin{equation}\label{orth}
    \frac{i}{\zeta^m}\left(\frac{\zeta-1}{\zeta} \right)\int \limits
    _{-\infty}^{\infty} d\tau\, \rho(\tau;\; \zeta)p_n(\tau;\; \zeta)p_m(\tau;\;
    \zeta)=\delta_{n\,m},
\end{equation}
with respect to the weight function,
\begin{equation}\label{rho}
 \rho(\tau;\; \zeta)=\frac{1}{2 \pi i}\Gamma\left(\frac{1}{2}+i\tau \right)
\Gamma\left(\frac{1}{2}-i\tau \right)
 (-\zeta)^{i\tau+\frac{1}{2}}.
\end{equation}
In (\ref{rho}) it is considered that $|\arg(-\zeta)|<\pi$. The
function $q_n^{(\pm)}$ is given by
\begin{equation}\label{qnpm}
 q_n^{(\pm)}(\tau; \; \zeta)=(-1)^n\frac{n!\Gamma\left(\frac{1}{2}\pm i\tau\right)}
 {\Gamma\left(n+\frac{3}{2}\pm i\tau\right)}\;
 {_2F_1\left(\frac{1}{2}\pm i\tau,\, n+1 ;\; n+\frac{3}{2}\pm i\tau;\;
 \zeta^{\mp 1}\right)}.
\end{equation}

Similarly, the nonzero elements of the tridiagonal matrix ${\bf
h}_{\eta}$ of the operator $\hat{\mathfrak{h}}_{\eta}$
(\ref{hxiheta}) in the basis $\left\{\psi_n(\eta)
\right\}_{n=0}^{\infty}$ (\ref{B1}) are
\begin{equation}\label{heta}
    h^{\eta}_{n,\,n}=b-i k+2 b n,\quad h^{\eta}_{n,\,n-1}=(b+i k)n,
    \quad h^{\eta}_{n,\,n+1}=(b-i k)(n+1).
\end{equation}
Obviously, the nonzero elements of the matrix ${\bf Q}_{\eta}$ of
the operator $\eta$ in the basis (\ref{B1}) are defined by
(\ref{Qxieta}). Thus, the elements of the matrix ${\bf
g}^{\eta(\pm)}$, which is inverse of the matrix $\left[{\bf
h}_{\eta}+2kt{\bf I}_{\eta}+\mu C {\bf Q}_{\eta} \right]$ (${\bf
I}_{\eta}$ is the unit matrix) of the one-dimensional operator
$\left[\hat{\mathfrak{h}}_{\eta}+2kt+\mu C \eta \right]$, are
expressed as
\begin{eqnarray}\label{gv_pm}
  g^{\eta(+)}_{n_1,\,n_2}(\tau;\; \gamma) &=& \frac{i}{2\gamma}
  \left(\frac{\zeta-1}{\zeta} \right)\frac{\lambda^{n_2-n_1}}{\zeta^{n_2}}\,
   p_{n_{<}}(\tau; \; \zeta)\, q_{n_{>}}^{(+)}(\tau; \; \zeta), \\
  g^{\eta(-)}_{n_1,\,n_2}(\tau;\; \gamma) &=& \frac{i}{2\gamma}
  \left(\frac{\zeta-1}{\zeta} \right)\frac{\lambda^{n_2-n_1}}{\zeta^{n_2}}\,
   p_{n_{<}}(\tau; \; \zeta)\,\zeta^{n_{>}+1}\, q_{n_{>}}^{(-)}(\tau; \; \zeta).
\end{eqnarray}

Note that $q_n^{(+)}$ $\left(q_n^{(-)}\right)$ are defined in the
upper (lower) half of the complex $\gamma$ plane where $|\zeta|>1$
($|\zeta|<1$). $q_n^{(+)}$ can be analytically continued onto the
lower half-plane by using the relation for hypergeometric functions
(15.3.7) in \cite{Abramowitz}, which transforms into the following
relationship between $q_n^{(+)}$ and $q_n^{(-)}$:
\begin{equation}\label{qpm}
    q_n^{(+)}(\tau; \; \zeta)=\zeta^{n+1}q_n^{(-)}(\tau; \;
    \zeta)+2\pi i \rho(\tau; \; \zeta)p_n(\tau; \; \zeta)
\end{equation}
Then the formulae for the analytical continuations of
$g^{\xi(+)}_{n_1,\,n_2}$ and $g^{\eta(+)}_{n_1,\,n_2}$ are
\begin{eqnarray}\label{acfxi}
  g^{\xi(+)}_{n_1,\,n_2}(\tau;\; \gamma) &=& g^{\xi(-)}_{n_1,\,n_2}(\tau;\; \gamma)
  -\frac{\pi}{\gamma}\left(\frac{\zeta-1}{\zeta} \right)
  \frac{\theta^{n_1}}{\lambda^{n_2}}\,
  \rho(\tau; \; \zeta)\,p_{n_1}(\tau; \; \zeta)\,p_{n_2}(\tau; \; \zeta),
  \\[3mm]
  \label{acfeta}
  g^{\eta(+)}_{n_1,\,n_2}(\tau;\; \gamma) &=& g^{\eta(-)}_{n_1,\,n_2}(\tau;\; \gamma)
  -\frac{\pi}{\gamma}\left(\frac{\zeta-1}{\zeta} \right)
  \frac{\theta^{n_2}}{\lambda^{n_1}}
  \rho(\tau; \; \zeta)\,p_{n_1}(\tau; \; \zeta)\,p_{n_2}(\tau; \; \zeta).
\end{eqnarray}

\subsection{Two-dimensional operators}
Note first that the completeness relations for eigenfunctions of the
one-dimensional operators $\left[\hat{\mathfrak{h}}_{\xi}+2kt+\mu C
\xi \right]$ and $\left[\hat{\mathfrak{h}}_{\eta}+2kt+\mu C \eta
\right]$ may take the form (see, e. g., \cite{JP1})
\begin{equation}\label{Cxieta}
    \pm\frac{\gamma}{i \pi}\int \limits _{-\infty}^{\infty}d\tau\,g_{n_1,\,n_2}^{\xi(\pm)}(\tau;\; \zeta)=
    \pm\frac{\gamma}{i \pi}\int \limits _{-\infty}^{\infty}d\tau\,g_{n_1,\,n_2}^{\eta(\pm)}(\tau;\; \zeta)=
    \frac{1}{2}\,\delta_{n_1,\,n_2}.
\end{equation}
The two-dimensional operator
$\left[\hat{\mathfrak{h}}_{\xi}+\hat{\mathfrak{h}}_{\eta}+2kt_0+\mu
C(\xi+\eta) \right]$ is also treated analytically within the context
of the basis $\left\{\phi_{n,\,m}(\xi,\,\eta)
\right\}_{n,\,m=0}^{\infty}$ (\ref{B2}). In particular, in view of
(\ref{Cxieta}), the inverse of the infinite matrix
\begin{equation}\label{h2mCQ}
\left[{\bf h}_{\xi}\otimes{\bf I}_{\eta}+{\bf I}_{\xi}\otimes{\bf
    h}_{\eta}+2kt_0{\bf I}_{\xi}\otimes{\bf I}_{\eta}+\mu C \left(
{\bf Q}_{\xi}\otimes{\bf I}_{\eta}+{\bf I}_{\xi}\otimes{\bf
Q}_{\eta}\right)\right]
\end{equation}
can be represented in the form of a convolution integral
\begin{equation}\label{Gpm}
  {\bf G}^{(\pm)}(t_0;\;\mathcal{E}) = \pm \frac{\gamma}{i \pi}\int
  \limits_{-\infty}^{\infty}d \tau \,{\bf g}^{\xi(\pm)}(\tau;\; \gamma)
  \otimes{\bf g}^{\eta(\pm)}(\tau_0-\tau;\; \gamma)
\end{equation}
where $\tau_0=\frac{k}{\gamma}\,t_0$. Finally we noted that the
matrix representation ${\bf G}^{(\pm)}$ (\ref{Gpm}) of the
two-dimensional Green's function operator is symmetric in $\xi$ and
$\eta$ unlike the corresponding formula obtained in \cite{JP2}.

\newpage
\begin{figure*}[ht]
\centerline{\psfig{figure=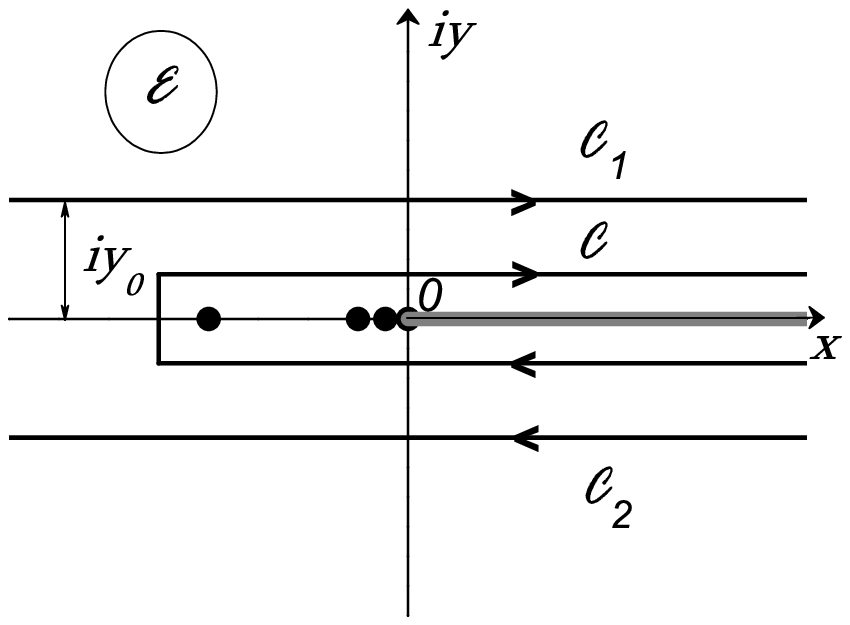,width=1\textwidth}} \caption{The
paths of integration on the physical energy sheet. The gray line is
the branch cut along the positive real axis. The poles of the
integrand in (\ref{CRS}) which occur at
$\mathcal{E}_{\ell}=-\frac{\left(kt\right)^2}{2\ell^2}$,
$\ell=1,2,\, \ldots$ for $t<0$ are shown as solid circles.}
\end{figure*}

\newpage
\begin{figure*}[ht]
\centerline{\psfig{figure=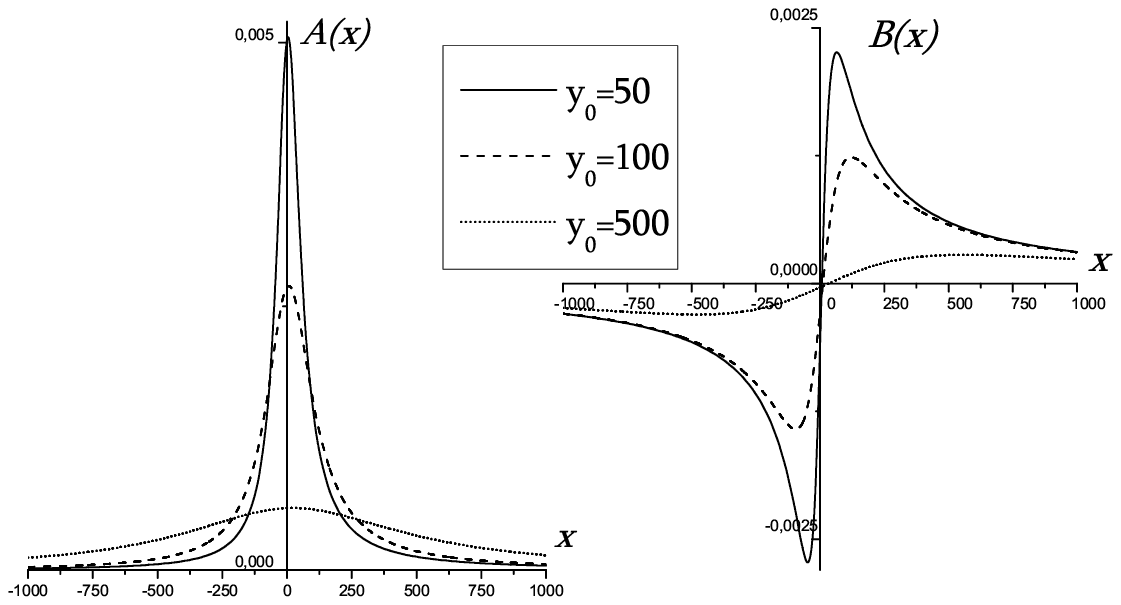,width=1.2\textwidth}} \caption{The
real $A(x)$ and imaginary $B(x)$ parts of $\frac{1}{2\pi
i}\,G^{(+)}_{0}\left(-\frac{2}{k};\;\mathcal{E}\right)$ for
$\mathcal{E}=x+iy_0$ with different $y_0$.}
\end{figure*}

\newpage
\begin{figure*}[ht]
\centerline{\psfig{figure=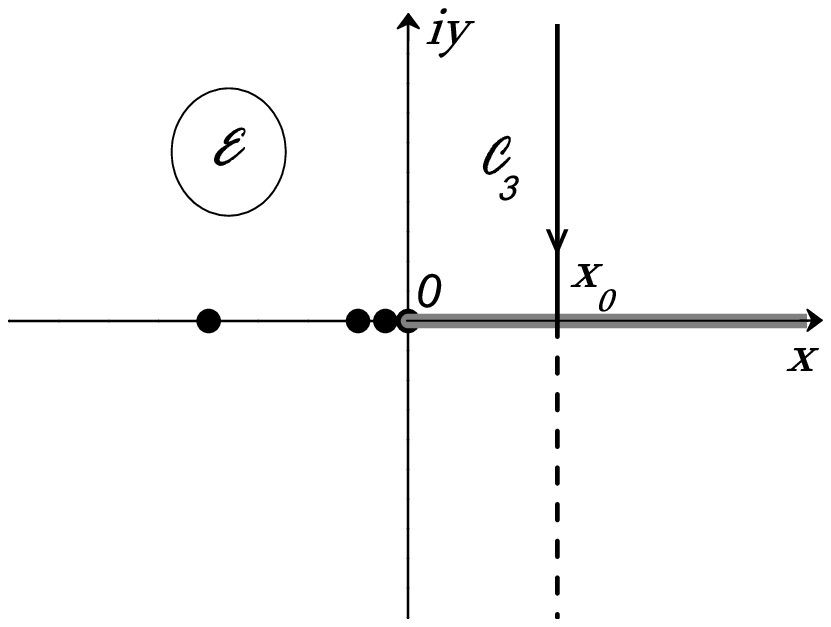,width=1.2\textwidth}} \caption{The
path of integration $\mathcal{C}_3$. The solid line is the part of
$\mathcal{C}_3$ which remains on the physical sheet. The part of
$\mathcal{C}_3$ which moves onto the unphysical sheet is depicted by
the dashed line.}
\end{figure*}

\newpage
\begin{table}
\caption{The numerical results obtained for integrals along the
contours $\mathcal{C}_1$ and $\mathcal{C}_3$ and for the element
$\left[\underline{\mathfrak{h}}\,\underline{\mathfrak{G}}^{(+)}\right]_{0,\,0}$
of the matrix product
$\underline{\mathfrak{h}}\underline{\mathfrak{G}}^{(+)}$.
}\label{T1}
\begin{ruledtabular}
\begin{tabular}{cc}
contour $\mathcal{C}_1$ & contour $\mathcal{C}_3$\\
\cline{1-1}\cline{2-2}$\mathcal{E}_1=iy_0+x_1$,
$\mathcal{E}_1=iy_0+x_1$,& $\mathcal{E}_1=\frac{k^2}{2}+iy_1$,
$\mathcal{E}_2=\frac{k^2}{2}+iy_2$,\\
$\mathcal{E}_3=k^2-iy_0-\frac{1}{2}\left(x_1+x_2 \right)$ &
$\mathcal{E}_3=\frac{k^2}{2}-\frac{i}{2}\left(y_1+y_2 \right)$\\
\hline
  $y_0=50\hphantom{0}$, $v^{(1)}_0=0.99996555-i\,2.7568672\times10^{-6}$&
            $v^{(3)}_0=1.0000132-i\,1.0613277\times10^{-4}$\\
  $y_0=100$, $v^{(1)}_0=0.99996555+i\,4.8873452\times10^{-6}$\\
  $y_0=500$, $v^{(1)}_0=0.99996539+i\,4.9623285\times10^{-6}$\\
$y_0=100$ & \\
\cline{1-1} &\\
           $ \begin{array}{l}
            w_1=-0.50009660-i\,0.75786560\times10^{-4}\\
            w_2=-0.50011010-i\,0.77244545\times10^{-4}\\
            w_3=\hphantom{-}1.00013285 -i\,1.3211843\times10^{-4}\\
\mathcal{I}_0=0.79996314\times10^{-9}-i\,0.62352239\times10^{-8}\\
 \end{array}$&
           $\begin{array}{l}
            w_1=1.49998186-i\,0.36607003\times10^{-3}\\
            w_2=1.49996754-i\,0.36818990\times10^{-3}\\
            w_3=0.99998324-i\,0.42641600\times10^{-3}\\
           \left[\underline{\mathfrak{h}}\underline{\mathfrak{G}}^{(+)}\right]_{0,\,0}
 =0.99999770 -i\,2.3034284\times10^{-5}
 \end{array}$\\
\end{tabular}
\end{ruledtabular}
\end{table}

\end{document}